\documentclass{emulateapj}

\usepackage{lscape}

\shorttitle{Spin Diagrams for bh binaries with aligned spins}
\shortauthors{Rezzolla et al.}

\begin{document}

\title{Spin Diagrams for equal-mass black-hole binaries with aligned spins}
\author{Luciano Rezzolla,\altaffilmark{1,3} 
  Ernst Nils Dorband,\altaffilmark{1} 
  Christian Reisswig,\altaffilmark{1}
  Peter Diener,\altaffilmark{2,3} 
  Denis Pollney,\altaffilmark{1} 
  Erik Schnetter,\altaffilmark{2,3} 
  B\'{e}la Szil\'{a}gyi\altaffilmark{1}} 
\altaffiltext{1}{Max-Planck-Institut f\"ur Gravitationsphysik,
Albert-Einstein-Institut, Potsdam-Golm, Germany}

\altaffiltext{2}{Center for Computation \& Technology, Louisiana State
University, Baton Rouge, LA, USA}

\altaffiltext{3}{Department of Physics and Astronomy, Louisiana State
University, Baton Rouge, LA, USA }

\begin{abstract}
  Binary black-hole systems with spins aligned with the orbital angular
  momentum are of special interest as they may be the preferred end-state
  of the inspiral of generic supermassive binary black-hole systems. In
  view of this, we have computed the inspiral and merger of a large set
  of binary systems of equal-mass black holes with spins aligned with the
  orbital angular momentum but otherwise arbitrary. By least-square
  fitting the results of these simulations we have constructed two ``spin
  diagrams'' which provide straightforward information about the recoil
  velocity $|v_{\rm kick}|$ and the final black-hole spin $a_{\rm fin}$
  in terms of the dimensionless spins $a_1$ and $a_2$ of the two initial
  black holes. Overall they suggest a maximum recoil velocity of $|v_{\rm
  kick}| = 441.94\,{\rm km/s} $, and minimum and maximum final spins
  $a_{\rm fin} = 0.3471$ and $a_{\rm fin} = 0.9591$, respectively.
\end{abstract}


\maketitle

\section{INTRODUCTION}
\label{intro}

A number of recent developments in numerical relativity have allowed
for stable evolution of binary black holes and opened the door to
extended and systematic studies of these systems. Of particular
interest to astrophysics are the calculations of the recoil velocity
and of the spin of the final black hole produced by the merger. It is
well known that a binary with unequal masses or spins will
radiate gravitational energy asymmetrically. This results in an uneven
flux of momentum, providing a net linear velocity to the final black
hole. The knowledge of both the ``kick'' velocity and of the final
spin could have a direct impact on studies of the evolution of
supermassive black holes and on statistical studies on the dynamics of
compact objects in dense stellar systems.

Over the past year, a number of simulations have been carried out to
determine the recoil velocities for a variety of binary black-hole
systems. Non-spinning but unequal-mass binaries were the first systems
to be studied and several works have now provided an accurate mapping
of the unequal-mass space of
parameters~\citep{Herrmann:2006ks,Baker:2006vn,Gonzalez:2006md}. More
recently, the recoils from binaries with spinning black holes have
also been considered by investigating equal-mass binaries in which the
spins of the black holes are either aligned with the orbital angular
momentum~\citep{Herrmann:2007ac,Koppitz-etal-2007aa}, or not. In the
first case, a systematic investigation has shown that the largest
recoil possible from such systems is of the order of
$450\,\mathrm{km/s}$~\citep{Pollney:2007ss}. In the second case,
instead, specific configurations with spins orthogonal to the orbital
one have been shown to lead to recoils as high as
$2500\,\mathrm{km/s}$~\citep{Campanelli:2007ew,Gonzalez:2007hi},
suggesting a maximum kick of about $4000\,\mathrm{km/s}$ for
maximally-spinning black holes~\citep{Campanelli:2007cg}. Recoil
velocities of this magnitude could lead to the ejection of massive
black holes from the hosting galaxies, with important consequences on
their cosmological evolution.

Here, we extend the analysis carried out in~\citet{Pollney:2007ss} of
binary black hole systems with equal-mass and spins aligned with the
orbital one. Our interest in this type of binaries stems from the fact
that systems of this type may represent a preferred end-state of the
binary evolution. Post-Newtonian studies have shown that in vacuum the
gravitational spin-orbit coupling has a tendency to align the spins
when they are initially close to the orbital
one~\citep{Schnittman:2004vq}. Furthermore, if the binary evolves in a
disc, as expected for supermassive black holes, the matter can exert a
torque tending to align the spins~\citep{Bogdanovic:2007hp}. Finally,
a recoiling supermassive black hole could retain the inner part of its
accretion disc and thus the fuel for a continuing QSO phase lasting
millions of years as it moves away from the galactic
nucleus~\citep{Loeb:2007}. Yet, the analysis of QSOs from the Sloan
Digital Sky Survey shows no evidence for black holes carrying an
accretion disc and hence for very large recoiling
velocities~\citep{Bonning:2007}.

\begin{table*}
\caption{\label{tableone}Binary sequences for which numerical
  simulations have been carried out, with different columns referring
  to the puncture initial location $\pm x/M$, the linear momenta $\pm
  p/M$, the mass parameters $m_i/M$, the dimensionless spins $a_i$,
  the normalized ADM mass ${\widetilde M_{_{\rm ADM}}}\equiv M_{_{\rm
  ADM}}/M$ measured at infinity, and the normalized ADM angular
  momentum ${\widetilde J_{_{\mathrm{ADM}}}}\equiv
  J_{_{\mathrm{ADM}}}/M^2$. Finally, the last six columns contain the
  numerical and fitted values for $|v_{\rm kick}|$ (in
  $\mathrm{km/s}$), $\,a_{\rm fin}$ and
  the corresponding errors.}
\vspace{0.1cm}
\begin{ruledtabular}
\begin{tabular}{|l|ccccccll|rrcccc|}
~					&
{$\pm x/M$} 				&
{$\pm p/M$} 				& 
{$m_1/M$} 				&
{$m_2/M$} 				& 
{$a_1$} 				&
{$a_2$} 				&
{${\widetilde M_{_{\rm ADM}}}$} 	&
{${\widetilde J_{_{\mathrm{ADM}}}}$}	&
{$|v_{_{\mathrm{kick}}}|$}	&
{$|v_{_{\mathrm{kick}}}^{\rm fit}|$}	&
{err.~($\%$)}				&
{$a_{_{\mathrm{fin}}}$}	&
{$a_{_{\mathrm{fin}}}^{\rm fit}$}	&
{err.~($\%$)}\\
\tableline
\hline
$r0$   & 3.0205 & 0.1366 & 0.4011 & 0.4009 & -0.584 & 0.584  & 0.9856 & 0.825  & 261.75 & 258.09 & 1.40  & 0.6891 & 0.6883 & 0.12 \\
$r1$   & 3.1264 & 0.1319 & 0.4380 & 0.4016 & -0.438 & 0.584  & 0.9855 & 0.861  & 221.38 & 219.04 & 1.06  & 0.7109 & 0.7105 & 0.06 \\ 
$r2$   & 3.2198 & 0.1281 & 0.4615 & 0.4022 & -0.292 & 0.584  & 0.9856 & 0.898  & 186.18 & 181.93 & 2.28  & 0.7314 & 0.7322 & 0.11 \\ 
$r3$   & 3.3190 & 0.1243 & 0.4749 & 0.4028 & -0.146 & 0.584  & 0.9857 & 0.935  & 144.02 & 146.75 & 1.90  & 0.7516 & 0.7536 & 0.27 \\  
$r4$   & 3.4100 & 0.1210 & 0.4796 & 0.4034 &  0.000 & 0.584  & 0.9859 & 0.971  & 106.11 & 113.52 & 6.98  & 0.7740 & 0.7747 & 0.08 \\  
$r5$   & 3.5063 & 0.1176 & 0.4761 & 0.4040 &  0.146 & 0.584  & 0.9862 & 1.007  & 81.42  & 82.23  & 1.00  & 0.7948 & 0.7953 & 0.06 \\  
$r6$   & 3.5988 & 0.1146 & 0.4638 & 0.4044 &  0.292 & 0.584  & 0.9864 & 1.044  & 45.90  & 52.88  & 15.21 & 0.8150 & 0.8156 & 0.07 \\  
$r7$   & 3.6841 & 0.1120 & 0.4412 & 0.4048 &  0.438 & 0.584  & 0.9867 & 1.081  & 20.59  & 25.47  & 23.70 & 0.8364 & 0.8355 & 0.11 \\  
$r8$   & 3.7705 & 0.1094 & 0.4052 & 0.4052 &  0.584 & 0.584  & 0.9872 & 1.117  & 0.00   & 0.00   & 0.00  & 0.8550 & 0.855  & 0.00 \\
\hline
$ra0$  & 2.9654 & 0.1391 & 0.4585 & 0.4584 & -0.300 & 0.300  & 0.9845 & 0.8250 & 131.34 & 132.58 & 0.95  & 0.6894 & 0.6883 & 0.16 \\  
$ra1$  & 3.0046 & 0.1373 & 0.4645 & 0.4587 & -0.250 & 0.300  & 0.9846 & 0.8376 & 118.10 & 120.28 & 1.85  & 0.6971 & 0.6959 & 0.17 \\  
$ra2$  & 3.0438 & 0.1355 & 0.4692 & 0.4591 & -0.200 & 0.300  & 0.9847 & 0.8499 & 106.33 & 108.21 & 1.77  & 0.7047 & 0.7035 & 0.17 \\  
$ra3$  & 3.0816 & 0.1339 & 0.4730 & 0.4594 & -0.150 & 0.300  & 0.9848 & 0.8628 & 94.98  & 96.36  & 1.46  & 0.7120 & 0.7111 & 0.13 \\  
$ra4$  & 3.1215 & 0.1321 & 0.4757 & 0.4597 & -0.100 & 0.300  & 0.9849 & 0.8747 & 84.74  & 84.75  & 0.01  & 0.7192 & 0.7185 & 0.09 \\
$ra6$  & 3.1988 & 0.1290 & 0.4782 & 0.4602 & 0.000  & 0.300  & 0.9850 & 0.9003 & 63.43  & 62.19  & 1.95  & 0.7331 & 0.7334 & 0.04 \\  
$ra8$  & 3.2705 & 0.1261 & 0.4768 & 0.4608 & 0.100  & 0.300  & 0.9852 & 0.9248 & 41.29  & 40.55  & 1.79  & 0.7471 & 0.7481 & 0.13 \\  
$ra10$ & 3.3434 & 0.1234 & 0.4714 & 0.4612 & 0.200  & 0.300  & 0.9853 & 0.9502 & 19.11  & 19.82  & 3.72  & 0.7618 & 0.7626 & 0.11 \\  
$ra12$ & 3.4120 & 0.1209 & 0.4617 & 0.4617 & 0.300  & 0.300  & 0.9855 & 0.9750 & 0.00   & 0.00   & 0.00  & 0.7772 & 0.7769 & 0.03 \\  
\hline
$s0$   & 2.9447 & 0.1401 & 0.4761 & 0.4761 & 0.000  & 0.000  & 0.9844 & 0.8251 & 0.00   & 0.00   & 0.00  & 0.6892 & 0.6883 & 0.13 \\  
$s1$   & 3.1106 & 0.1326 & 0.4756 & 0.4756 & 0.100  & 0.100  & 0.9848 & 0.8749 & 0.00   & 0.00   & 0.00  & 0.7192 & 0.7185 & 0.09 \\  
$s2$   & 3.2718 & 0.1261 & 0.4709 & 0.4709 & 0.200  & 0.200  & 0.9851 & 0.9251 & 0.00   & 0.00   & 0.00  & 0.7471 & 0.7481 & 0.13 \\  
$s3$   & 3.4098 & 0.1210 & 0.4617 & 0.4617 & 0.300  & 0.300  & 0.9855 & 0.9751 & 0.00   & 0.00   & 0.00  & 0.7772 & 0.7769 & 0.03 \\  
$s4$   & 3.5521 & 0.1161 & 0.4476 & 0.4476 & 0.400  & 0.400  & 0.9859 & 1.0250 & 0.00   & 0.00   & 0.00  & 0.8077 & 0.8051 & 0.33 \\  
$s5$   & 3.6721 & 0.1123 & 0.4276 & 0.4276 & 0.500  & 0.500  & 0.9865 & 1.0748 & 0.00   & 0.00   & 0.00  & 0.8340 & 0.8325 & 0.18 \\  
$s6$   & 3.7896 & 0.1088 & 0.4002 & 0.4002 & 0.600  & 0.600  & 0.9874 & 1.1246 & 0.00   & 0.00   & 0.00  & 0.8583 & 0.8592 & 0.11 \\  
\hline
$t0$   & 4.1910 & 0.1074 & 0.4066 & 0.4064 & -0.584 & 0.584  & 0.9889 & 0.9002 & 259.49 & 258.09 & 0.54  & 0.6868 & 0.6883 & 0.22 \\  
$t1$   & 4.0812 & 0.1103 & 0.4062 & 0.4426 & -0.584 & 0.438  & 0.9884 & 0.8638 & 238.37 & 232.62 & 2.41  & 0.6640 & 0.6658 & 0.27 \\  
$t2$   & 3.9767 & 0.1131 & 0.4057 & 0.4652 & -0.584 & 0.292  & 0.9881 & 0.8265 & 200.25 & 205.21 & 2.48  & 0.6400 & 0.6429 & 0.45 \\  
$t3$   & 3.8632 & 0.1165 & 0.4053 & 0.4775 & -0.584 & 0.146  & 0.9879 & 0.7906 & 174.58 & 175.86 & 0.73  & 0.6180 & 0.6196 & 0.26 \\  
$t4$   & 3.7387 & 0.1204 & 0.4047 & 0.4810 & -0.584 & 0.000  & 0.9878 & 0.7543 & 142.62 & 144.57 & 1.37  & 0.5965 & 0.5959 & 0.09 \\  
$t5$   & 3.6102 & 0.1246 & 0.4041 & 0.4761 & -0.584 & -0.146 & 0.9876 & 0.7172 & 106.36 & 111.34 & 4.68  & 0.5738 & 0.5719 & 0.33 \\  
$t6$   & 3.4765 & 0.1294 & 0.4033 & 0.4625 & -0.584 & -0.292 & 0.9874 & 0.6807 & 71.35  & 76.17  & 6.75  & 0.5493 & 0.5475 & 0.32 \\  
$t7$   & 3.3391 & 0.1348 & 0.4025 & 0.4387 & -0.584 & -0.438 & 0.9873 & 0.6447 & 35.36  & 39.05  & 10.45 & 0.5233 & 0.5227 & 0.11 \\  
$t8$   & 3.1712 & 0.1419 & 0.4015 & 0.4015 & -0.584 & -0.584 & 0.9875 & 0.6080 & 0.00   & 0.00   & 0.00  & 0.4955 & 0.4976 & 0.42 \\  
\hline
$u1$   & 2.9500 & 0.1398 & 0.4683 & 0.4685 & -0.200 & 0.200  & 0.9845 & 0.8248 & 87.34  & 88.39  & 1.20  & 0.6893 & 0.6883 & 0.15 \\  
$u2$   & 2.9800 & 0.1384 & 0.4436 & 0.4438 & -0.400 & 0.400  & 0.9846 & 0.8249 & 175.39 & 176.78 & 0.79  & 0.6895 & 0.6883 & 0.17 \\  
$u3$   & 3.0500 & 0.1355 & 0.3951 & 0.3953 & -0.600 & 0.600  & 0.9847 & 0.8266 & 266.39 & 265.16 & 0.46  & 0.6884 & 0.6883 & 0.01 \\  
$u4$   & 3.1500 & 0.1310 & 0.2968 & 0.2970 & -0.800 & 0.800  & 0.9850 & 0.8253 & 356.87 & 353.55 & 0.93  & 0.6884 & 0.6883 & 0.01 \\  
\hline
\end{tabular} \\
\end{ruledtabular}
\end{table*}

\section{NUMERICAL SETUP AND INITIAL DATA}
\label{numerical simulations}

The numerical simulations have been carried out using the CCATIE code, a
three-dimensional finite-differences code using the Cactus Computational
Toolkit~\citep{cactusweb} and Carpet mesh refinement
infrastructure~\citep{schnetter_etal:2004}. The main features of the code
have been recently reviewed in~\citet{Pollney:2007ss}, where the code has
been employed using the so-called ``moving-punctures''
technique~\citep{Baker:2006yw, Campanelli:2005dd}. The initial data
consists of five sequences with constant orbital angular momentum, which
is however different from sequence to sequence. In the $r$ and
$ra$-sequences, the initial spin of one of the black holes ${\mathbf
S}_2$ is held fixed along the $z$-axis and the spin of the other black
hole is varied so that the spin ratio $a_1/a_2$ takes the values between
$-1$ and $+1$, with $a_i\equiv{\mathbf S}_i/M_i^2$. In the $t$-sequence,
instead, the spin with a negative $z$-component is held fixed, while in
the $s$ and $u$-sequences $a_1/a_2=1$ and $-1$, respectively.  In all
cases, the masses are $M_i = M/2=1/2$. For the orbital initial data
parameters we use the effective-potential method, which allows one to
choose the initial data parameters such that the resulting physical
parameters (\textit{e.g.,} masses and spins) describe a binary black-hole
system on a quasi-circular orbit. The free parameters are: the coordinate
locations ${\mathbf C}_i$, the mass parameters $m_i$, the linear momenta
${\mathbf p}_i$, and the spins ${\mathbf S}_i$. Quasi-circular orbits are
then selected by setting ${\mathbf p}_1 = -{\mathbf p}_2$ to be
orthogonal to ${\mathbf C}_2 - {\mathbf C}_1$, so that ${\mathbf L}
\equiv {\mathbf C}_1 \times {\mathbf p}_1 + {\mathbf C}_2 \times {\mathbf
p}_2$ is the orbital angular momentum. The initial parameters are
collected in the left part of Table~\ref{tableone}, while the right part
reports the results of simulations. For all of them we have employed 8
levels of refinement and a minimum resolution $0.024\,M$, which has been
reduced to $0.018\,M$ for binaries $r5,\,r6$. Note that our results for
the $u$-sequence differ slightly from those reported
by~\citet{Herrmann:2007ac}, probably because of our accounting of the
integration constant in $|v_{\rm kick}|$~\citep{Pollney:2007ss}.

\begin{figure}
\begin{center} 
\scalebox{0.45}{\includegraphics{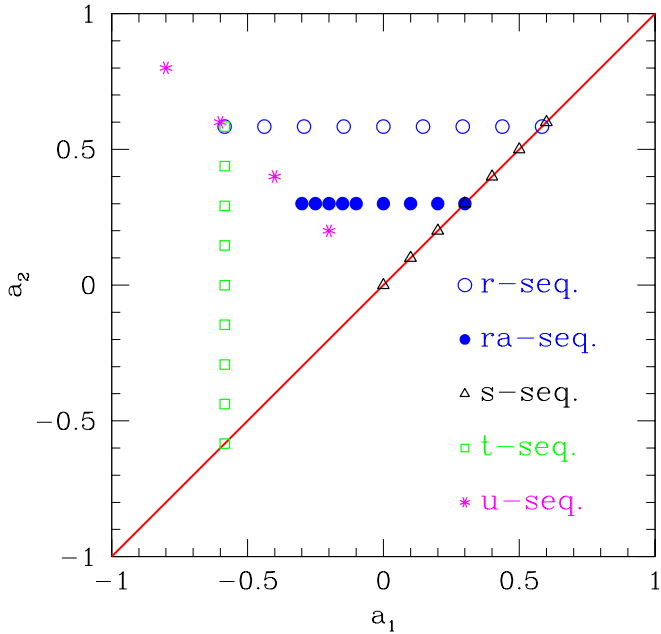}}
\caption{\label{plotone} Position in the $(a_1,\,a_2)$ space of the five
  sequences $r,\, ra,\, s,\, t$, and $u$ for which the inspiral and
  merger has been computed.}
\end{center}
\end{figure}

\section{SPIN DIAGRAMS AND FITS}
\label{spin_diagrams}

Clearly, the recoil velocity and the spin of the final black hole are
among the most important pieces of information to be extracted from the
inspiral and coalescence of binary black holes. For binaries with equal
masses and aligned but otherwise arbitrary spins, this information
depends uniquely on the dimensionless spins of the two black holes
$a_1,\,a_2$ and can therefore be summarized in the portion of the
$(a_1,\,a_2)$ plane in which the two spins vary. It is therefore
convenient to think in terms of ``spin diagrams'', which summarize in a
simple way all of the relevant information. In addition, since the
labelling ``$1$'' and ``$2$'' is arbitrary, the line $a_1=a_2$ in the
spin diagram has important symmetries: the recoil velocity vector
undergoes a $\pi$-rotation, \textit{i.e.}, ${\vec v}_{\rm kick}
(a_1,\,a_2) = - {\vec v}_{\rm kick} (a_2,\,a_1)$ but $|v_{\rm kick}
(a_1,\,a_2)| = |v_{\rm kick} (a_2,\,a_1)|$, while no change is expected
for the final spin, \textit{i.e.}, $a_{\rm fin}(a_1,\,a_2) = a_{\rm
fin}(a_2,\,a_1)$. These symmetries not only allow us to consider only one
portion of the $(a_1,\,a_2)$ space (\textit{cf.}  Fig.~\ref{plotone}),
thus halving the computational costs (or doubling the statistical
sample), but they will also be exploited later on to improve our
fits. The position of the five sequences within the $(a_1,\,a_2)$ space
is shown in Fig.~\ref{plotone}.

Overall, the data sample computed numerically consists of 38 values
for $|v_{\rm kick}|$ and for $a_{\rm fin}$ which, for simplicity, we
have considered to have constant error-bars of $8\ {\rm km/s}$ and
$0.01$, which represent, respectively, the largest errors reported
in~\citet{Pollney:2007ss}. In both cases we have modelled the data
with generic quadratic functions in $a_1$ and $a_2$ so that, in the
case of the recoil velocity, the fitting function is
\begin{equation}
|v_{\rm kick}| = | c_0         + c_1 a_1 + c_2 a^2_1 +
	           d_0 a_1 a_2 + d_1 a_2 + d_2 a^2_2 |\,.
\label{vk_1}
\end{equation}
Note that the fitting function on the right-hand-side of (\ref{vk_1})
is smooth everywhere but that its absolute value is not smooth along
the diagonal $a_1=a_2$. Using~(\ref{vk_1}) and a blind least-square
fit of the data, we obtained the coefficients (in $\mathrm{km/s}$)
\begin{eqnarray}
& c_0 =    0.67 \pm  1.12 \,, \quad  
& d_0 =  -18.56 \pm  5.34 \,, 
\nonumber \\
& c_1 = -212.85 \pm  2.96 \,, \quad  
& d_1 =  213.69 \pm  3.57 \,, 
\nonumber \\
& c_2 =   50.85 \pm  3.48 \,, 
& d_2 =  -40.99 \pm  4.25 \,,
\label{vk_fit_1}
\end{eqnarray}        
with a reduced-$\chi^2 = 0.09$. Clearly, the errors in the
coefficients can be extremely large and this is simply the result of
small-number statistics. However, the fit can be improved by
exploiting some knowledge about the physics of the process to simplify
the fitting expressions. In particular, we can use the constraint that
no recoil velocity should be produced for binaries having the same
spin, \textit{i.e.}, that $|v_{\rm kick}|=0$ for $a_1=a_2$, or the
symmetry condition across the line $a_1=a_2$. Enforcing both
constraints yields
\begin{equation}
c_0 =  0\,,   \quad
c_1 = -d_1\,, \quad 
c_2 = -d_2\,, \quad
d_0 = 0\,,
\label{vk_coeff_2}
\end{equation}
thus reducing the fitting function~(\ref{vk_1}) to the simpler expression
\begin{equation}
|v_{\rm kick}| = |c_1 (a_1 - a_2) + c_2 (a_1^{\,2} - a_2^{\,2})|\,.
\label{vk_3}
\end{equation}

\begin{figure}
\begin{center} 
\scalebox{0.45}{\includegraphics{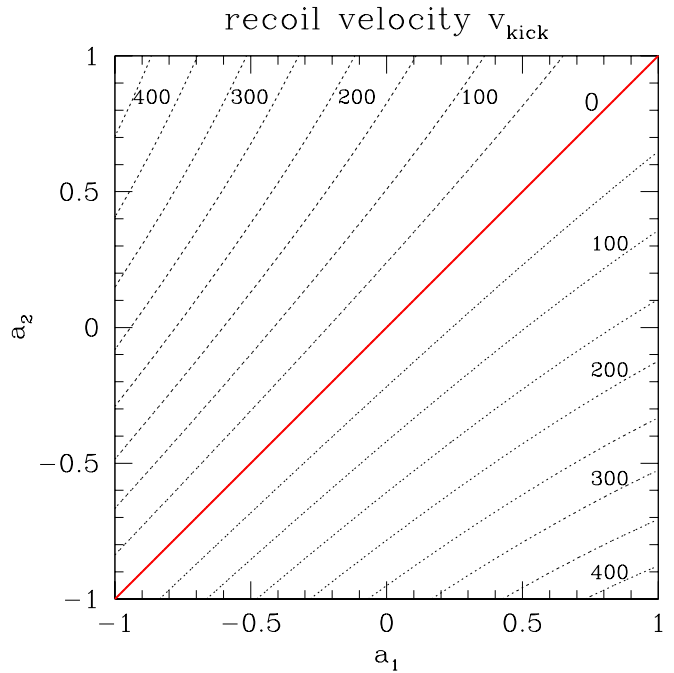}}
\caption{\label{plottwo} Contour plots of $|v_{\rm kick}|$ as a
  function of the spin parameters $a_1$ and $a_2$. The diagram has
  been computed using expressions~(\ref{vk_3}) and~(\ref{vk_fit_2}).}
\end{center}
\end{figure}

Performing a least-square fit using (\ref{vk_3}) we then obtain
\begin{eqnarray}
&  c_1 = -220.97 \pm 0.78\,, \quad
&  c_2 =   45.52 \pm 2.99\,, \quad       
\label{vk_fit_2}
\end{eqnarray}
with a comparable reduced-$\chi^2 = 0.14$, but with error-bars that
are much smaller on average. Because of this, we consider
expression~(\ref{vk_3}) as the best description of the data at
second-order in the spin parameters. Using~(\ref{vk_3}) and
~(\ref{vk_fit_2}), we have built the contour plots shown in
Fig.~\ref{plottwo}.

A few remarks are worth making. Firstly, we recall that post-Newtonian
calculations have so far derived only the linear contribution in the
spin to the recoil velocity (see~\citet{Favata_etal:2004} and
references therein). However, the size of the quadratic coefficient
(\ref{vk_fit_2}) is not small when compared to the linear one and it
can lead to rather sizeable corrections. These are maximized when
$a_1=0$ and $a_2=\pm 1$, or when $a_1=\pm 1$ and $a_2=0$, and can be
as large as $\sim 20\%$; while these corrections are smaller than
those induced by asymmetries in the mass, they are instructive in
pointing out the relative importance of spin-spin and spin-orbit
effects during the merger and can be used as a guide in further
refinements of the post-Newtonian treatments. Secondly, expression
(\ref{vk_3}) clearly suggests that the maximum recoil velocity should
be found when the asymmetry is the largest and the spins are
antiparallel, \textit{i.e.}, $a_1=-a_2$. Thirdly, when $a_2={\rm
const.}$, expression (\ref{vk_3}) confirms the quadratic scaling
proposed in~\citet{Pollney:2007ss} with a smaller data set
[\textit{cf.,} eq.~(42) there]. Fourthly, for $a_1=-a_2$,
expression~(\ref{vk_3}) is only linear and reproduces the scaling
suggested by~\citet{Herrmann:2007ac}. Finally, using~(\ref{vk_3}) the
maximum recoil velocity is found to be $|v_{\rm kick}| = 441.94 \pm
1.56\ {\rm km/s}$, in very good agreement with the results
of~\citet{Herrmann:2007ac} and~\citet{Pollney:2007ss}.

In the same way we have first fitted the data for $a_{\rm fin}$, with
a function
\begin{equation}
a_{\rm fin} = p_0         + p_1 a_1 + p_2 a^2_1 +
	      q_0 a_1 a_2 + q_1 a_2 + q_2 a^2_2\,,
\label{af_1}
\end{equation}
and found coefficients with very large error-bars. As a result, also
for $a_{\rm fin}$ we resort to physical considerations to constrain
the coefficients $p_0 \ldots q_2$. More specifically, we expect that, at
least at lowest order, binaries with equal and opposite spins will not
contribute to the final spin and thus behave essentially as
nonspinning binaries. Stated differently, we assume that $a_{\rm fin}
= p_0$ for binaries with $a_1=-a_2$. In addition, enforcing the
symmetry condition across the line $a_1=a_2$ we obtain
\begin{equation}
p_1 =  q_1 \,, \quad 
p_2 =  q_2 = q_0/2\,,
\label{af_coeff}
\end{equation}
so that the fitting function~(\ref{af_1}) effectively reduces to 
\begin{equation}
a_{\rm fin} = p_0 + p_1 (a_1 + a_2) + p_2 (a_1 + a_2)^2\,.
\label{af_2}
\end{equation}
Performing a least-square fit using (\ref{af_2}) we then obtain
\begin{eqnarray}
&  p_0 =  0.6883 \pm 0.0003 \,,\quad
&  p_1 =  0.1530 \pm 0.0004 \,, 
\nonumber \\ 
&  p_2 = -0.0088 \pm 0.0005 \,,
\label{af_fit_2}
\end{eqnarray}
with a reduced-$\chi^2=0.02$. 

\begin{figure}
\begin{center} 
\scalebox{0.45}{\includegraphics{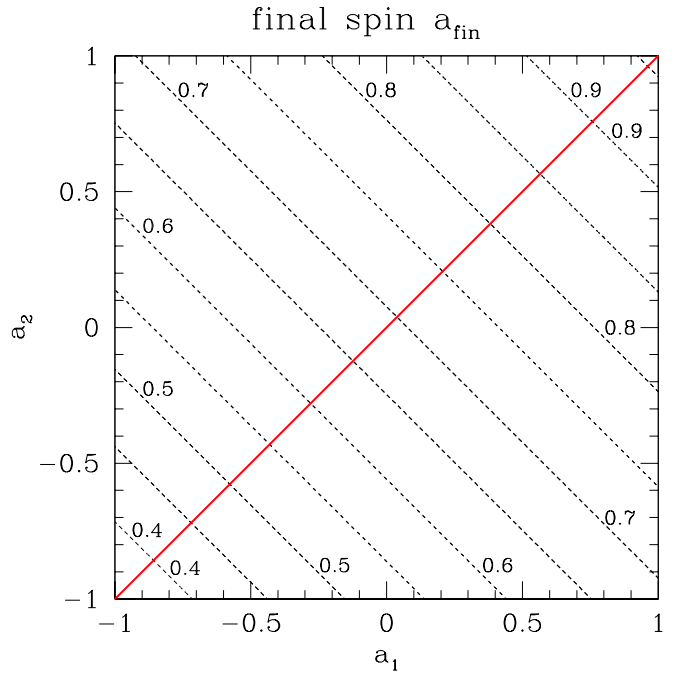}}
\caption{\label{plotthree} Contour plots of $a_{\rm fin}$ as a function of
  the spin parameters $a_1$ and $a_2$. The diagram has been computed
  using expressions~(\ref{af_2}) and~(\ref{af_fit_2}).}
\end{center}
\end{figure}

It should be noted that the coefficient of the quadratic term
in~(\ref{af_fit_2}) is much smaller then the linear one and with much
larger error-bars. Given the small statistics it is hard to assess
whether a quadratic dependence is necessary or if a linear one is the
correct one (however, see also the comment below on a possible
interpretation of expression~(\ref{af_2})). In view of this, we have
repeated the least-square fit of the data enforcing the
conditions~(\ref{af_coeff}) together with $p_2=0$
(\textit{i.e.},~adopting a linear fitting function) and obtained $p_0
= 0.6855 \pm 0.0007$ and $p_1 = 0.1518 \pm 0.0012$, with a worse
reduced-$\chi^2=0.16$. Because the coefficients of the lowest-order
terms are so similar, both the linear and the quadratic fits are well
within the error-bars of the numerical simulations. Nevertheless,
since a quadratic scaling yields smaller residuals, we consider it to be the
best representation of the data and have therefore computed the
contour plots in Fig.~\ref{plotthree} using~(\ref{af_2})
and~(\ref{af_fit_2}).

Here too, a few remarks are worth making: Firstly, the fitted value
for the coefficient $p_0$ agrees very well with the values reported by
several groups~\citep{Gonzalez:2006md,berti_etal:2007} when studying
the inspiral of unequal-mass nonspinning binaries. Secondly,
expression~(\ref{af_2}) has maximum values for $a_1=a_2$, suggesting
that the maximum and minimum spins are $a_{\rm fin} = 0.9591 \pm
0.0022$ and $a_{\rm fin} = 0.3471 \pm 0.0224$, respectively. Thirdly,
the quadratic scaling for $a_{\rm fin}$ substantially confirms the
suggestions of~\citet{Campanelli:2006gf} but provides more accurate
coefficients. Finally, although very simple,
expression~(\ref{af_fit_2}) lends itself to an interesting
interpretation. Being effectively a power series in terms of the
initial spins of the two black holes, its zeroth-order term can be
seen as the orbital angular momentum not radiated in gravitational
waves and which amounts, at most, to $\sim 70\%$ of the final
spin. The first-order term, on the other hand, can be seen as the
contribution to the final spin coming from the initial spins of the
two black holes and this contribution, together with the one coming
from the spin-orbit coupling, amounts at most to $\sim 30\%$ of the
final spin. Finally, the second-order term, which is natural to expect
as nonzero in this view, can then be related to the spin-spin
coupling, with a contribution to the final spin which is of $\sim 4\%$
at most.
			   
As a side remark we also note that the monotonic behaviour expressed
by (\ref{af_fit_2}) does not show the presence of a local maximum of
$a_{\rm fin} \simeq 0.87$ for $a_1 = a_2 \sim 0.34$ as
suggested by~\citet{Damour:2001tu} in the effective one-body (EOB)
approximation. Because the latter has been shown to be in good
agreement with numerical-relativity simulations of nonspinning black
holes~\citep{Damour_Nagar:2007, Damour_etal:2007}, additional
simulations will be necessary to refute these results or to improve
the EOB approximation for spinning black holes.

Reported in the right part of Table~\ref{tableone} are also the fitted
values for $a_{\rm fin}$ and $|v_{\rm kick}|$ obtained through the
fitting functions~(\ref{vk_3}) and~(\ref{af_2}), and the corresponding
errors.  The latter are of few percent for most of the cases and
increase up to $\sim 20\%$ only for those binaries with very small
kicks and which are intrinsically more difficult to calculate. As a
concluding remark we note that the fitting coefficients computed here
have been constructed using overall moderate values of the initial
spin; the only exception is the binary $u4$ which has the largest spin
and which is nevertheless fitted with very small errors
(\textit{cf.}~Table~\ref{tableone}). In addition, since the submission
of this work, another group has reported results from equal-mass
binaries with spins as high as $a_1=a_2=\pm
0.9$~\citep{Marronetti_etal:2007}.  Although also for these very
high-spin binaries the error in the predicted values is of $1\%$ at
most, a larger sample of high-spin binaries is necessary to validate
that the fitting expressions~(\ref{vk_3}) and~(\ref{af_2}) are robust
also at very large spins.

\section{CONCLUSIONS}
\label{CONCLUSIONS}

We have performed least-square fits to a large set of
numerical-relativity data. These fits, combined with symmetry
arguments, yield analytic expressions for the recoil velocity and
final black hole spin resulting from the inspiral and merger of
equal-mass black holes whose spins are parallel or antiparallel to the
orbital angular momentum.  Such configurations represent a small
portion of the space of parameters, but may be the preferred ones if
torques are present during the evolution.  Using the analytic
expressions we have constructed two spin diagrams that summarize
simply this information and predict a maximum recoil velocity of
$|v_{\rm kick}| = 441.94 \pm 1.56\ {\rm km/s} $ for systems with
$a_1=-a_2=1$ and maximum (minimum) final spin $a_{\rm fin} = 0.9591
\pm 0.0022\, (0.3471 \pm 0.0224)$ for systems with $a_1=a_2=1\,
(-1)$. 

\vspace{0.25cm}\noindent It is a pleasure to thank Thibault Damour and
Alessandro Nagar for interesting discussions. The computations were
performed on the supercomputing clusters of the AEI. This work was
supported in part by the DFG grant SFB/Transregio~7.

\bigskip
\bigskip

\noindent{NOTE ADDED IN PROOF.} Since the publication on the preprint
archive of this analysis, our work on the modelling of the final spin has
progressed rapidly, yielding new results that complement and complete the
ones presented here. In particular, the work published
in~\citet{Rezzolla:2008a} complements the analysis carried here to
unequal-mass, equal-spin aligned binaries, while the work reported
in~\citet{Rezzolla:2008b} extends it to generic binaries.

\end{document}